\documentclass[11pt]{article}
\usepackage{amsmath,amsfonts}
\usepackage{graphicx}
\usepackage[numbers]{natbib}
\usepackage{algorithm}
\usepackage{amsmath}
\usepackage{amssymb,amsfonts,amscd,amsxtra,bm}
\usepackage{bm}
\usepackage{mathrsfs}
 
\usepackage{graphicx}
\usepackage{textcomp}
\usepackage{xcolor}

\usepackage{float}
\usepackage{stfloats}
\usepackage{cuted}
\usepackage{subfig}
\usepackage{multicol}
\usepackage{multirow}
\usepackage{tikz}
\usetikzlibrary{shapes.geometric, arrows}
\usepackage{enumerate} 
\tikzstyle{startstop} = [rectangle, rounded corners, 
minimum width=3cm, 
minimum height=1cm,
text centered, 
draw=white, 
fill=red!30]

\tikzstyle{io} = [trapezium, 
trapezium stretches=true,  
trapezium left angle=70, 
trapezium right angle=110, 
minimum width=3cm, 
minimum height=1cm, text centered, 
draw=white, fill=blue!30]

\tikzstyle{process} = [rectangle, 
minimum width=3cm, 
minimum height=1cm, 
text centered, 
text width=3cm, 
draw=black, 
fill=white!30]

\tikzstyle{decision} = [diamond, 
minimum width=3cm, 
minimum height=1cm, 
text centered, 
draw=black, 
fill=white!30]
\tikzstyle{arrow} = [thick,->,>=stealth]

\usepackage{mathtools}
\usepackage{subfig}
\usepackage{caption}
\usepackage{array}
\usepackage[noend]{algpseudocode}
\usepackage{amssymb,graphicx}
\usepackage{mathtools}
\usepackage{mathptmx}
\usepackage[affil-it]{authblk}
\usepackage{amsthm}
\usepackage{amssymb}
\usepackage{mathrsfs}   % For \mathscr (script-like calligraphy)
\usepackage{bm} 
\usepackage{booktabs}
\usepackage{multirow,multicol}
\usepackage{hyperref}
\usepackage{adjustbox}
\usepackage{url}
\usepackage{physics} 
\usepackage[margin=1in]{geometry}

\title{Parametric Sensitivity Analysis: Local and Global Approaches in Stochastic Biochemical Models}

\author{
 Kannon Hossain\textsuperscript{1}, Roger Sidje \textsuperscript{2}, Fahad Mostafa\textsuperscript{3}\\

\textsuperscript{1} Department of Mathematics, Florida Gulf Coast University, Fort Myers, FL 33965, USA\par
\vspace{1 mm}
\textsuperscript{2} Department of Mathematics, The University of Alabama, Tuscaloosa, AL 35487, USA\par
\vspace{1 mm}
\textsuperscript{3} School of Mathematical and Natural Sciences, Arizona State University, Tempe, AZ 85281,
USA\\
}
\date{}

\begin{document}

\maketitle

\begin{abstract}
The recent advancements in mathematical modeling of biochemical systems have generated increased interest in sensitivity analysis methodologies. There are two primary approaches for analyzing these mathematical models: the stochastic approach, which employs chemical master equations (CME), and the deterministic approach, which utilizes ordinary differential equations (ODEs). The intractable discrete states present in most biochemical processes render the direct simulation of the CME infeasible. Moment closure approximations are recognized for their numerical efficiency in estimating the statistics of the CME solution. Since classical sensitivity analysis is not directly applicable to stochastic modeling, this work conducts sensitivity analysis using moment-based ordinary differential equations (ODEs) to identify key parameters that significantly influence the dynamics of the model. We conduct numerical tests to evaluate the effectiveness of both local and global sensitivity analyses of the moment-based ODEs. These tests enable us to examine how variations in input parameters influence the model's output.
\end{abstract}

 \section{Introduction}
 The significance of quantitative mathematical modeling of biochemical reaction networks has increased considerably due to the developments in measurement technology in biochemical processes\cite{newman2003structure,barabasi2004network,bertaux2014modeling,neuert2013systematic}. Chemical reactions at the cellular level with low copy numbers of molecules reacting with different species are known to be stochastic\cite{elowitz2002stochastic}. The chemical master equation (CME), which governs the time evolution of the probability distribution of the continuous-time Markov chain (CTMC)\cite{goutsias2013markovian}, is commonly used for modeling biochemical reaction networks due to their ability to capture the stochasticity\cite{gillespie1992rigorous}. Because of the curse of dimensionality, the size of CME can be extremely large or theoretically infinite, making it often not directly solvable even for the simplest systems\cite{wolf2010solving,munsky2006finite}. Several direct approximation techniques, including both exact and inexact methods \cite{munsky2006finite,burrage2006krylov,hegland2007solver,wolf2010solving,kazeev2014direct,vo2017adaptive,dinh2020adaptive,gillespie1977exact,van1992stochastic,gillespie2000chemical,cao2006efficient}, have been implemented to find the approximate solution of the CME but are numerically expensive. A computationally cost-effective method has been developed to approximate the CME solution, which describes the probability distribution in terms of its moments\cite{barzel2012stochastic,singh2010stochastic,gomez2007mass,hespanha2008moment,lee2009moment,lee2013moment,ale2013general,hossain2025study,hossain2024case,hossain2023parameter,2023parameter,hossain2024bayesian}. The time evolution of moments that are derived from the CME presented by ordinary differential equations (ODEs) requires approximation for their solution. These approximations are known as moment closures and exist in numerous forms depending on their underlying probability distributions\cite{hespanha2008moment}.\\
Since many biochemical processes rely on some unknown fluctuating ranges of parameters, it is therefore vital to analyze the influence of the model parameters on the dynamics of the system, often known as sensitivity analysis\cite{rabitz1983sensitivity}.  Sensitivity analysis is applicable in various aspects of stochastic biochemical models, such as model simplification, experimental analysis, model reduction, and parameter estimation problems.\cite{varma1999parametric,qian2020sensitivity}. A model is considered sensitive if small changes in the parameter cause significant changes in the system output, whereas no significant changes in output indicate that the model is robust \cite{zi2011sensitivity,gunawan2005sensitivity}. There are two main types of sensitivity analysis: local and global approaches. Local sensitivity analysis\cite{kirch2016effect} is a traditional approach that examines the effects of small perturbations on model outputs, whereas global sensitivity analysis\cite{sobol2001global} demonstrates the impact of large variations in the model parameters on the model outputs. Although local sensitivity approaches are easier to execute, global sensitivity analysis techniques are more widely used since their outcomes are independent of core variables.\\
In this article, our comprehensive work for sensitivity analysis has the following structure: We first use zero-moment approximation to find the moment-based ODEs for each molecular species in the model. We then conduct local sensitivity analysis using parameter perturbation as well as by computing normalized sensitivity, while global sensitivity analysis is examined using Sobol' method. We evaluate the quality and efficiency of all the methods by considering two systems biology examples, namely the birth-death process and the dimerization process.\\
We organize the rest of the paper as follows: Section ~\ref{sec:back} provides methodological backgrounds on the chemical master equation, method of moments, local and global sensitivity analysis. In section ~\ref{sec:Numerical}, we present the numerical tests. Finally, closing remarks on section ~\ref{sec:Conclusion}.
 
 \section{Background}\label{sec:back}
 \subsection{The Chemical Master Equation} 
 Consider a chemical reaction system consisting of $N$ molecular species that interact through $M$ reactions of the form
 \begin{equation}
 	R_k: a_{1 k} S_1+\cdots+a_{N k} S_N \stackrel{c_k}{\longrightarrow} b_{1 k} S_1+\cdots+b_{N k} S_N
 \end{equation}
 where $a_{ik}$ and $b_{ik}$ are coefficients denoting the number of $S_i$ molecules, with $c_k$ being the reaction rate constant for $1 \le k \le M$ and $1 \le i \le N$. We denote $\boldsymbol{x}(t)=\left(x_1, \ldots, x_N\right)^T$ as the state of the system at time $t$. The propensity function $\alpha_k(\boldsymbol{x}(t))$ of reaction $R_k$ at the current state $\boldsymbol{x}(t)$ is defined so that the probability of such a reaction occurring during the infinitesimal time interval $[t, t+d t)$ is $\alpha_k(\boldsymbol{x}(t)) d t$. When reaction $R_k$ happens, the state vector is updated with the stoichiometric vector $\boldsymbol{\nu}_k$, representing the change in species numbers.\\
 Denote $P(\boldsymbol{x}, t)=\operatorname{Prob}\{\boldsymbol{x}(t)=\boldsymbol{x}\}$, the probability that the system is at state $\boldsymbol{x}$ at time $t$. As given in~\cite{gillespie1992rigorous}, a characterization CME is that
 \begin{equation} \label{eq:CME}
 	\begin{aligned}
 \frac{d P(\boldsymbol{x}, t)}{d t}=\sum_{k=1}^M \alpha_k\left(\boldsymbol{x}-\boldsymbol{\nu}_k\right) P\left(\boldsymbol{x} - \boldsymbol{\nu}_k, t\right)  	
 -  \sum_{k=1}^M \alpha_k(\boldsymbol{x}) P(\boldsymbol{x}, t) 
 	\end{aligned}
 \end{equation}
Let $\boldsymbol{X}$ be the set of all possible states, if we order these states as $\boldsymbol{X} = \{\boldsymbol{x}_1, \ldots, \boldsymbol{x}_n\}$, where $\boldsymbol{x}_i=\left(x_{1 i}, \ldots, x_{N i}\right)^T$ and $n$ is the total number of states, then~\eqref{eq:CME} defines a set of ODEs  
	\begin{equation}
		\label{eq:CMEode}
		\dot{{\bm{p}}}\left(t\right)={\bm{A}}\cdot {\bm{p}}\left(t\right),  \qquad t\in[0,t_f]
	\end{equation} 
The transition rate matrix $\boldsymbol{A}=\left[a_{i j}\right] \in \mathbb{R}^{n \times n}$ is defined as
	$$
	a_{i j}=\left\{\begin{array}{l}
		-\sum_{k=1}^M \alpha_k\left(\boldsymbol{x}_j\right), \quad \text { if } i=j \\
		\alpha_k\left(\boldsymbol{x}_j\right), \quad \text { if } \boldsymbol{x}_i=\boldsymbol{x}_j+\boldsymbol{\nu}_k \\
		0, \text { otherwise }
	\end{array} .\right.
	$$
	From~\eqref{eq:CMEode} the probability vector at the end point $t_f$ is
	\begin{equation}
		\label{eq:sol_cme}
		\boldsymbol{p}\left(t_f\right)=\exp \left(t_f \boldsymbol{A}\right)\cdot \boldsymbol{p}(0)
\end{equation}
 
 \subsection{Method of Moments}\label{sec:mom}
 We denote $\mathbb{E}\left[x_i\right] = \mu_i$ is the first moment of the $i$th species. To obtain it, we multiply Eq.~\eqref{eq:CME} by $x_i$ on both sides, and summing over all reachable states $\boldsymbol{x} = \left(x_1, \ldots, x_N\right)^T$,
 \begin{equation}\label{eq:first}
 	\begin{aligned}
 		\sum_{\boldsymbol{x}\in \boldsymbol{X}} x_i \frac{d P(\boldsymbol{x}, t)}{d t}=\sum_{\boldsymbol{x}\in \boldsymbol{X}} \left(\sum_{k=1}^M x_i \alpha_k (\boldsymbol{x}-\boldsymbol{\nu}_k) P(\boldsymbol{
 			x}-\boldsymbol{\nu}_k, t)
 		- x_i P(\boldsymbol{x}, t) \sum_{k=1}^M \alpha_k(\boldsymbol{x})\right)
 	\end{aligned}
 \end{equation}
 Similarly, for the second central moment, we multiply Eq.~\eqref{eq:CME} by $\left(x_i-\mu_i\right)\left(x_j-\mu_j\right)$ on both sides, and summing over all reachable states $\boldsymbol{x} = \left(x_1, \ldots, x_N\right)^T$,
 \begin{equation}\label{eq:second}
 	\begin{aligned}
 		\sum_{\boldsymbol{x}\in \boldsymbol{X}} \left(x_i-\mu_i\right)\left(x_j-\mu_j\right) \frac{d P(\boldsymbol{x}, t)}{d t}=  
 		\sum_{\boldsymbol{x}\in \boldsymbol{X}} \left(\sum_{k=1}^M \left(x_i-\mu_i\right)\left(x_j-\mu_j\right) \alpha_k (\boldsymbol{x}-\boldsymbol{\nu}_k) P(\boldsymbol{
 			x}-\boldsymbol{\nu}_k, t) \right. \\
 		\left.
 		-\left(x_i-\mu_i\right)\left(x_j-\mu_j\right) P(\boldsymbol{x}, t) \sum_{k=1}^M \alpha_k(\boldsymbol{x})\right)
 	\end{aligned}
 \end{equation}
 Following the derivation given in \cite{ lee2009moment}, where we apply a transformation $\boldsymbol{x}-\boldsymbol{\nu}_k \rightarrow \boldsymbol{x}$ using the fact that $\mathbb{E}\left[\alpha_k(\boldsymbol{x})\right]= \sum_{\boldsymbol{x}\in \boldsymbol{X}} \alpha_k(\boldsymbol{x}) P(\boldsymbol{x}, t)$ 
 with $\alpha_k(\boldsymbol{x})$ representing the $k^{\text {th }}$ reaction propensity at state $\boldsymbol{x}$, Eqs.~\eqref{eq:first} and ~\eqref{eq:second} can be written as
 \begin{equation}
 	\label{eq:CME_MOM_initiial}
 	\begin{aligned}
 		\frac{d \mathbb{E}\left[x_i\right]}{d t} = \sum_{k=1}^M \nu_{k, i} \mathbb{E}\left[\alpha_k(\boldsymbol{x})\right]
 \end{aligned}
 \end{equation} 
 \begin{equation} 
 	\label{eq:CME_MOM_initiial2}
 	\begin{aligned}
 		\frac{d \mathbb{E}\left[\left(x_i-\mu_i\right)\left(x_j-\mu_j\right)\right]}{d t}= 
 		\sum_{k=1}^M \biggl(\nu_{k, i} \mathbb{E} [(x_j-\mu_j) \alpha_k(\boldsymbol{x})]
 		+ \nu_{k, j} \mathbb{E}\left[\left(x_i-\mu_i\right) \alpha_k(\boldsymbol{x})\right] 
 		+ \nu_{k, i} \nu_{k, j} \mathbb{E}\left[\alpha_k(\boldsymbol{x})\right]\biggr)
 	\end{aligned}
 \end{equation} 
 If we denote $\mathbb{E}\left[x_i\right]= \mu_i$ and ${\mathbb{E}\left[\left(x_i-\mu_i\right)\left(x_j-\mu_j\right)\right]}= \sigma_{ij}$, then by applying multivariate Taylor series expansion of $\mathbb{E}\left[\alpha_k(\boldsymbol{x})\right]$ around the mean $\boldsymbol{\mu}=
 (\mu_1,\ldots,\mu_N)^T= \left(\mathbb{E}\left[x_1\right], \ldots, \mathbb{E}\left[x_N\right]\right)^T$, one can simplify Eqs.~\eqref{eq:CME_MOM_initiial} and ~\eqref{eq:CME_MOM_initiial2} as
 \begin{equation}
 	\label{eq:CME_MOM_initiial3}
 	\begin{aligned}
 		\frac{d \mu_i}{d t} =\sum_k \nu_{k, i}\left(a_k(\boldsymbol{\mu})+\frac{1}{2} \sum_{l, m} \frac{\partial^2 a_k(\boldsymbol{\mu})}{\partial x_l \partial x_m} \sigma_{l m}+R_3\right)
 	\end{aligned}
 \end{equation}
 \begin{equation}
 	\label{eq:CME_MOM_initiial4}
 	\begin{aligned}
 		\frac{d \sigma_{i j}}{d t} =\sum_k\left[\nu_{k, i}\left(\sum_l \frac{\partial a_k(\boldsymbol{\mu})}{\partial x_l} \sigma_{j l}+\bar{R}_3\right)  
 		+ \nu_{k, j}\left(\sum_l \frac{\partial a_k(\boldsymbol{\mu})}{\partial x_l} \sigma_{i l}+\bar{R}_3\right) 
 		+  \right. \\ \left.
 		\nu_{k, i} \nu_{k, j}\left(a_k(\boldsymbol{\mu})+\frac{1}{2} \sum_{l, m} \frac{\partial^2 a_k(\boldsymbol{\mu})}{\partial x_l \partial x_m} \sigma_{l m}+R_3\right)\right] 
 	\end{aligned}
 \end{equation}
 where $\mu_{i}$ is the first moment and $\sigma_{i j}$ is the  second central moment. The terms $\bar{R}_3$ and ${R}_3$ are related to the third central moment. In general, the lower-order $m^{th}$ moment depends on the higher-order $(m+1)^{th}$ moment, which forms an infinite series of ODEs that are usually intractable and cannot be solved analytically or numerically. A moment closure truncates the infinite ODEs to a finite one by eliminating the dependence of lower-order moments on higher-order moments. The system is considered closed by making assumptions about the underlying distribution, such as zero-moment approximation\cite{lakatos2015multivariate}, normal moment-approximation\cite{whittle1957use}, log-normal moment approximation\cite{whittle1957use}, and so on. The zero-moment approximation is a basic moment approximation method where the Taylor series in Eqs. ~\eqref{eq:CME_MOM_initiial3} and ~\eqref{eq:CME_MOM_initiial4} is truncated at a specific order, as if all subsequent higher-order moments were set to be zero.  In this paper, we will focus on the zero-moment approximation or zero closure, where the highest moment order is considered to be two and assumes that the covariance is zero between interacting species. By setting the third moment equal to zero in Eqs. ~\eqref{eq:CME_MOM_initiial3} and ~\eqref{eq:CME_MOM_initiial4}, we obtain the following closed systems:
 \begin{equation}
 	\label{eq:CME_MOM_initiial5}
 	\begin{aligned}
 		\frac{d \mu_i}{d t} =\sum_k \nu_{k, i}\left(a_k(\boldsymbol{\mu})+\frac{1}{2} \sum_{l, m} \frac{\partial^2 a_k(\boldsymbol{\mu})}{\partial x_l \partial x_m} \sigma_{l m}\right)
 	\end{aligned}
 \end{equation}
 \begin{equation} 
 	\label{eq:CME_MOM_initiial6}
 	\begin{aligned}
 		\frac{d \sigma_{i j}}{d t} =\sum_k\left[\nu_{k, i}\left(\sum_l \frac{\partial a_k(\boldsymbol{\mu})}{\partial x_l} \sigma_{j l} \right) 
 		+ \nu_{k, j}\left(\sum_l \frac{\partial a_k(\boldsymbol{\mu})}{\partial x_l} \sigma_{i l}\right)  
 		+ \nu_{k, i} \nu_{k, j}\left(a_k(\boldsymbol{\mu})+\frac{1}{2} \sum_{l, m} \frac{\partial^2 a_k(\boldsymbol{\mu})}{\partial x_l \partial x_m} \sigma_{l m}\right)\right]
 	\end{aligned}
 \end{equation} 
 Later on, we will use Eqs. ~\eqref{eq:CME_MOM_initiial5} and ~\eqref{eq:CME_MOM_initiial6} to derive the moment-based ODEs for each species of our test models. It is important to note that some people refer to the first moment as mean of the species and the second central moment as variance of the species. A detailed derivation of moment closure approximations can be found in these articles\cite{engblom2006computing,lee2013moment,lee2009moment}.

 \subsection{Local Sensitivity Analysis}\label{sec:sens}
  To outline the concepts, we consider the following ODEs
  \begin{equation}
  	\begin{aligned}
  		&\frac{d{\boldsymbol{y}}}{d t}=\boldsymbol{f}(\boldsymbol{y}, \boldsymbol{\theta}, t) \quad \forall {\boldsymbol{f}} \in {\mathbb R}^{n}, {\boldsymbol{y}} \in {\mathbb R}^{n}, \boldsymbol{\theta} \in {\mathbb R}^{m}\\
  		& {\boldsymbol{y}}\left(t_0\right)={\boldsymbol{y}}_0 \\
  	\end{aligned}
  \end{equation}
  where $\boldsymbol{f}$ is a function giving us the right-hand side of the ODEs, ${\boldsymbol{y}}$ is the vector of dependent variables, ${\boldsymbol{y}}_0$ is the vector of initial conditions, $t$ is the time and $\boldsymbol{\theta}$ is the vector of constant parameters. A straightforward and simplest method to calculate the local sensitivity is by computing the derivatives of the system output with respect to the input parameters\cite{tomovic1972general,varma1999parametric,griewank2008evaluating}. Mathematically, the sensitivity coefficients are defined as first-order partial derivatives of the model output ${\boldsymbol{y}}( t,{\bm{\theta}})$ with respect to the parameter ${\bm{\theta}}$,
  \begin{equation}\label{eq:fd}
  	S_{\bm{\theta}}={\frac{\partial \boldsymbol{y}}{\partial {\bm{\theta}}}} =\lim _{{\Delta {\bm{\theta}}} \to 0} \frac{\boldsymbol{y}({\bm{\theta}}+\Delta {\bm{\theta}})-\boldsymbol{y}({\bm{\theta}})}{\Delta {\bm{\theta}}}
  \end{equation}
Several methods\cite{rabitz1983sensitivity,qian2020sensitivity} exist for calculating the local sensitivities as well as computing the derivative in Eq.~\eqref{eq:fd}, such as the finite difference approximations, direct differentials method, adjoint method, automatic differentiations, and complex perturbation method. Unlike the direct differentials and other derivative-based methods, the computation of finite difference approximations does not depend on the dimensionality of the parameter space or the size of the state variable and thus allows for a faster calculation of the sensitivity coefficients. The finite difference method \cite{haftka1989recent,iott1985selecting} calculates the sensitivities using the forward differences approximations and is written as
  \begin{equation}\label{eq:sens1}
  	\begin{aligned}
  		S_{\bm{\theta}}={\frac{\partial \boldsymbol{y}}{\partial {\bm{\theta}}}} \approx \frac{\boldsymbol{y}({\bm{\theta}}+\Delta {\bm{\theta}})-\boldsymbol{y}({\bm{\theta}})}{\Delta {\bm{\theta}}}
  	\end{aligned}
  \end{equation}
In practice, the model parameters and state variables may have different units or orders of magnitude. As a result, it is more convenient to work with the relative sensitivity function (RSF), which is also known as normalized sensitivity coefficients and can be expressed as 
  \begin{equation}\label{eq:sens}
  	{\mathscr{S}}_{\bm{\theta}}= S_{\bm{\theta}}\cdot {\frac{\omega_{\bm{\theta}}}{\omega_{\boldsymbol{y}}}}
  \end{equation}
  where $\omega_{\boldsymbol{y}}$ is the scaling of variable $\boldsymbol{y}$ and $\omega_{\bm{\theta}}$ is the scaling of parameter ${\bm{\theta}}$. According to Kirch et al., \cite{kirch2016effect}, the normalized local sensitivity coefficients are invariant with respect to the scaling of model parameters and output variables. 
 
 \subsection{Global Sensitivity Analysis}
 \subsubsection{Sobol' Method}\label{sec:sobol}
The Sobol' method is a variance-based method that does not rely on any assumptions about the relationship between the inputs and outputs of a model. To summarize and simplify notation, we describe the following formulas derived in\cite{sobol2001global,saltelli2008global}, where $Y$ denotes a particular scalar output variable at a particular time and ${\bm{\mathscr{X}}}=(\mathscr{X}_1, \ldots, \mathscr{X}_k)$ is the set of input model parameters, expressing the system as 
  \begin{equation}
  	Y=f\left(\mathscr{X}_1, \ldots, \mathscr{X}_k\right) 
  \end{equation}
  The fundamental concept of the Sobol’ is to decompose the function $f\left(\mathscr{X}_1, \ldots, \mathscr{X}_k\right)$ into terms of increasing dimensionality, namely
  \begin{equation}\label{eq:sobol}
  	\begin{aligned}
  		f\left(\mathscr{X}_1, \ldots, \mathscr{X}_k\right)=  f_0+\sum_{i=1}^k f_i\left(\mathscr{X}_i\right)+\sum_{i=1}^k \sum_{j=i+1}^k f_{i j}\left(\mathscr{X}_i, \mathscr{X}_j\right)
  		+  \cdots 
  		+f_{1 \ldots k}\left(\mathscr{X}_1, \mathscr{X}_2, \ldots \mathscr{X}_k\right)
  	\end{aligned}
  \end{equation}
  The term $f_0$ is a constant, $f_i(\mathscr{X}_i)$ is called the $i$-th first order effect and $f_{i j}(\mathscr{X}_i, \mathscr{X}_j)$ the second order interaction between $\mathscr{X}_i$ and $\mathscr{X}_j$. Higher order interactions are defined analogously using other index sets $I \subseteq\{1, \ldots, k\}$. If the input parameters are mutually independent, then there exists a unique decomposition of Eq.~\eqref{eq:sobol} such that all the summands are mutually orthogonal. The variance of the output variable $Y$ then decomposed as 
  \begin{equation}\label{eq:sobol1}
  	\begin{aligned}
  		D=Var(Y)=Var\left(f\left(\mathscr{X}_1, \ldots, \mathscr{X}_k\right)\right)=  Var\left(f_0\right) + 
  		Var \left(\sum_{i=1}^k f_i\left(\mathscr{X}_i\right)\right)\\
  		+ Var\left(\sum_{i=1}^k \sum_{j=i+1}^k f_{i j}\left(\mathscr{X}_i, \mathscr{X}_j\right)\right)
  		+ \cdots 
  		+Var\left(f_{1 \ldots k}\left(\mathscr{X}_1,\mathscr{X}_2, \ldots \mathscr{X}_k\right)\right)
  	\end{aligned}
  \end{equation}
  Usually the Sobol' sensitivity indices are expressed in terms of conditional variances\cite{jansen1999analysis,oakley2004probabilistic,saltelli2010variance}. The first-order Sobol' indices or main effect measure the direct effect of each input parameter to the output variance. The first-order Sobol' indices for input parameter $\mathscr{X}_i$ is given by
  \begin{equation}
  	S_i=\frac{Var\left(E\left[Y \mid \mathscr{X}_i\right]\right)}{Var(Y)}
  \end{equation}
  where $E$ is the expected value. Here, $E\left[Y \mid \mathscr{X}_i\right]$ denotes the expected value of the output $Y$ when parameter $\mathscr{X}_i$ is fixed. $Var(Y)$ is the total variance of the output $Y$. The first-order Sobol' indices tell us the expected reduction in the variance of the model when we fix parameter $\mathscr{X}_i$. The sum of the first-order Sobol sensitivity indices can not exceed one.\\
 Homma and Saltelli\cite{homma1996importance} proposed the total effect sensitivity indices, which account for both the direct effect of parameter $\mathscr{X}_i$ and the effects arising from its interactions (covariance) with other parameters. The total Sobol' indices for input parameter $\mathscr{X}_i$ is defined as
  \begin{equation}
  	S_{T_i}= 1 - \frac{Var\left(E\left[Y \mid \mathscr{X}_{\sim i}\right]\right)}{Var(Y)}
  \end{equation}
  where $\mathscr{X}_{\sim i}$ denotes all model parameters except $\mathscr{X}_i$. In detail, the total effect sensitivity indices\cite{zi2011sensitivity} can be described as 
  \begin{equation}
  	S_{T_i}=S_i+S_{u i}
  \end{equation}
  where $S_{u i}$ is the complementary set of practical sensitivity indices that parameter $\mathscr{X}_i$ appears. For example, $S_{T_1}=S_1+$ $S_{12}+S_{13}+S_{123}$ is the total effect index of model parameter $\mathscr{X}_1$ for three-model input parameters. The total effect sensitivity index quantifies the overall effects of a parameter, in combination with any other parameter(s), on the model output\cite{zi2011sensitivity}. Sobol indices range from $0$ (not sensitive) to $1$ (highly sensitive). A value approaching 1 signifies that the associated input variable has a greater impact on the variance of the model output, meaning it is more sensitive. Sobol' introduced the Monte Carlo-based method to estimate the first and total order sensitivity indices. The numerical methodology employed to approximate the values of $S_i$ and $S_{T_i}$ can be found in\cite{saltelli2008global,saltelli2010variance}. The computation of Sobol' indices involves two primary steps. The first step involves drawing sample values for the input parameters. Several methods  \cite{rabitz1983sensitivity,qian2020sensitivity,zi2011sensitivity,gunawan2005sensitivity,varma1999parametric}, including Monte Carlo sampling, random sampling, the grid method, latin hypercube, Sobol' sequence, the winding stairs method, pseudo-random method, quasi-random sequences, and their adaptive variants exist for sampling purposes. In the second step, the sensitivity indices $S_i$ and $S_{T_i}$ are calculated\cite{sobol2001global,saltelli2008global,saltelli2010variance} using the samples obtained from the sampling method. Jansen\cite{jansen1999analysis} and Martinez\cite{martinez2011analyse} have developed methods for estimating the Sobol sensitivity indices, which are based on Monte Carlo simulation. Our experiment focuses on the Martinez methods, which are numerically stable and have been shown to be the most efficient\cite{saltelli2010variance}.
 
    \begin{figure}
 	\begin{center}
 		\includegraphics[width=.75\linewidth]{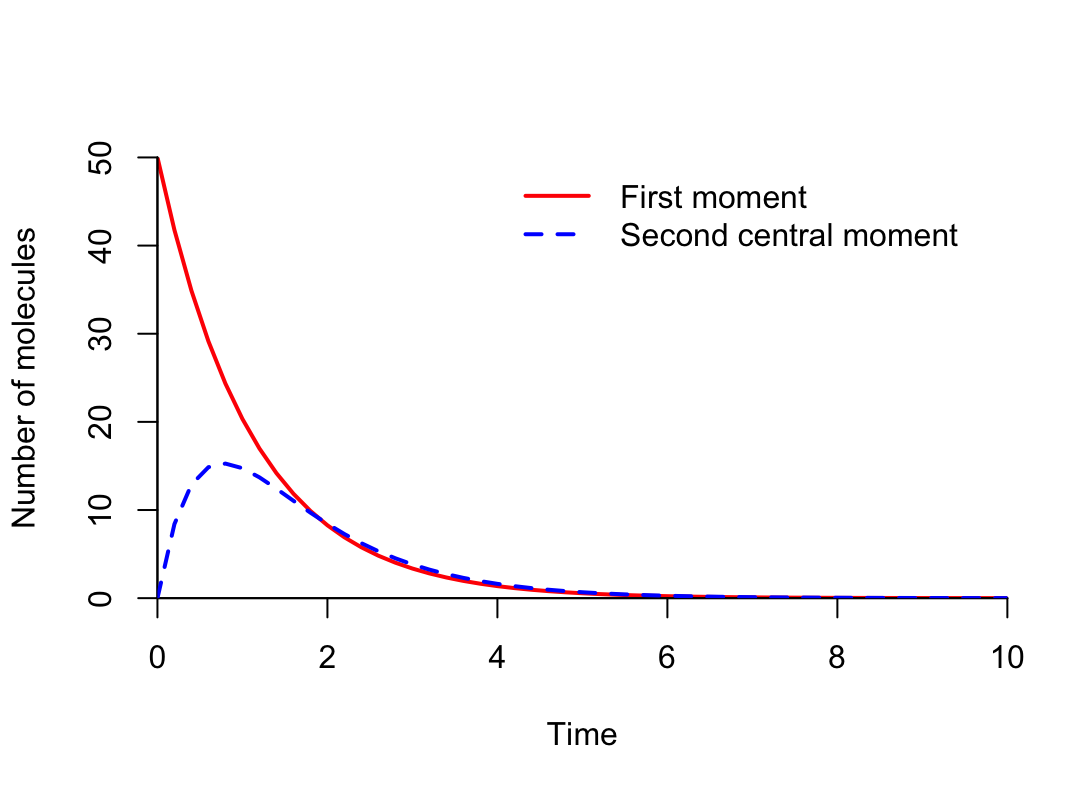}
 		\caption{First moment and second central moment of the birth-death process.}
 		\label{fig:BD}
 	\end{center}
 \end{figure}  
 
 \section{Numerical Tests}
 \label{sec:Numerical}
 \subsection{Birth-Death Process}
 Consider the birth-death process for the production and degradation of a protein species $X$ with a molecule $x$, given by the following reactions:
 $$
 \begin{array}{cc} 
 	R_1: & X \stackrel{c_1}{\longrightarrow} 2 X \\ [10 pt]
 	R_2: & X \stackrel{c_2}{\longrightarrow} \varnothing
 \end{array}
 $$
 where in reaction $R_{1}$ protein is produced at a constant rate $c_1$ and in reaction $R_{2}$ protein is degraded at a constant rate $c_2$. The propensity functions for $R_1$ and $R_2$ are given by $\alpha_1 = c_1 x$ and $\alpha_2 = c_2 x$, respectively, where the associated stoichiometric matrix is $S = (1\quad-1)$. By using Eq.~\eqref{eq:CME} one can derive the CME for the birth-death process\cite[p.~54]{gillespie2009moment} as follows:
 \begin{equation}
 	\begin{aligned}
 		\label{eq:bd_CME}
 		\frac{d P (x, t)}{d t}=c_1(x-1) P(x-1, t) - \left(c_1+c_2\right) x P(x, t)
 		+ c_2(x+1) P(x+1, t)
 	\end{aligned}
 \end{equation}
 Solving Eq.~\eqref{eq:bd_CME} would fully characterize the stochastic dynamics of the system. However, due to the curse of dimensionality, finding the solution for the whole probability distribution is computationally infeasible. Instead of simulating the entire probability distribution, a computationally cheap and more viable approach would be to find the first moment and the second central moment, which provides a decent description for the stochastic dynamics, such as the averaging behavior and evolution of the noise in the system. Finding the moment equations for systems involving non-linear reactions is harder than finding the moment equations for linear systems. For the non-linear case, moment equations for ODEs describing time evolution are not closed and depend on higher-order moments. As we previously discussed in this paper, we will only consider the highest moment order as two and assume the covariance between the reacting species is zero. Eqs.~\eqref{eq:CME_MOM_initiial5} and ~\eqref{eq:CME_MOM_initiial6} can be used to derive the following moment-based ODEs for birth-death process\cite[Supporting Information S1 p.~2]{kugler2012moment}
 \begin{equation}
 	\label{eq:BD_mean_differ}
 	\begin{aligned}
 		\frac{d \mu_1(t)}{d t}=\left(c_1-c_2\right) \mu_1(t),  \mu_1(0)=x_0 
 	\end{aligned}
 \end{equation}
 \begin{equation}
 	\label{eq:BD_var_differ}
 	\begin{aligned}
 		\frac{d \sigma_{11} (t)}{d t}=2\left(c_1-c_2\right) \sigma_{11}(t)+\left(c_1+c_2\right) \mu_1(t), \sigma_{11}(0)=0  
 	\end{aligned}
 \end{equation}
 where $\mu_1$ is the first moment and $\sigma_{11}$ is the second central moment of protein. The solutions of moment Eqs. ~\ref{eq:BD_mean_differ} and ~\ref{eq:BD_var_differ} are depicted in Fig.~\ref{fig:BD} with the known (true) parameter values: ${\bm{\theta}_{true}}= (c_1, c_2) = (0.10, 1.0)$ and initial conditions $(\mu_1(0), \sigma_{11}(0)) = (50, 0)$\cite{wilkinson2009stochastic} for the time $t=10s$.  
 
  \begingroup
 \setlength{\tabcolsep}{9pt} % Default value: 6pt
 \renewcommand{\arraystretch}{1.2} 
 \begin{table}[h]
 	\caption{Bounds on the support of the uniform distribution of the parameters in the birth-death process.}
 	\begin{center}
 		\begin{tabular}{@{}c c c @{}} 
 			\hline
 			parameter & lower bound & upper bound  \\
 			\hline 
 			$c_{1}$  & $0.05$ & $1.0$ \\
 			$c_{2}$  & $0.50$ & $2.0$ \\
 			\hline
 		\end{tabular}
 	\end{center}
 	\label{table:BD_bound}
 \end{table}
 \endgroup 
 
  \begin{figure}[!htp]
 	\begin{center}
 		\includegraphics[width=.75\linewidth]{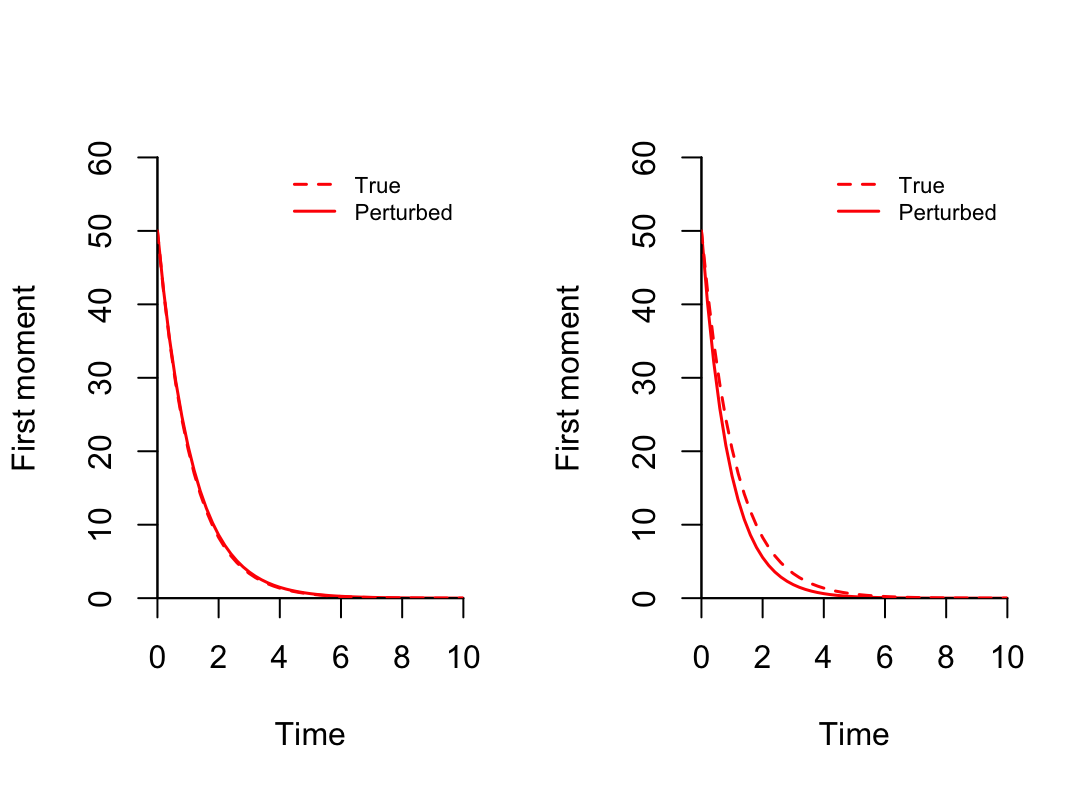}
 		\caption{The left plot shows the perturbation effect of $c_1$ when $c_2$ is fixed, while the right plot shows the perturbation effect of $c_2$ when $c_1$ is fixed for the first moment of the birth-death process.}
 		\label{fig:BD_local1}
 	\end{center}
 \end{figure}  
 
 \begin{figure}[!htp]
 	\begin{center}
 		\includegraphics[width=.75\linewidth]{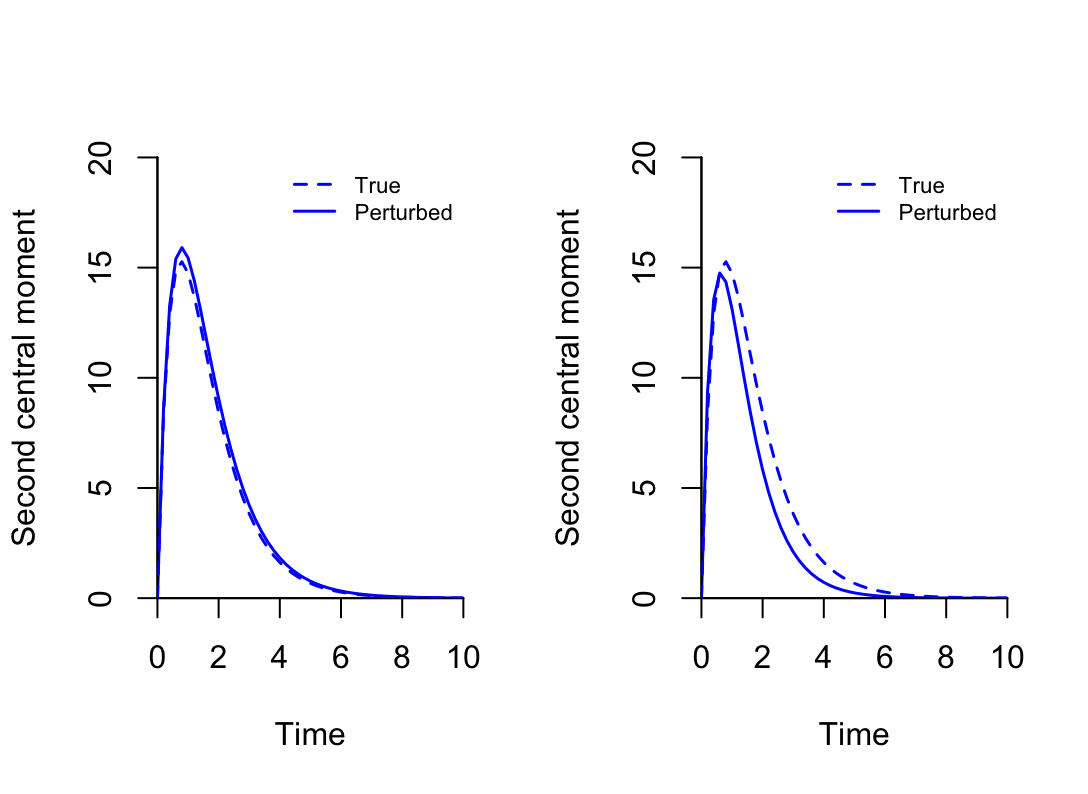}
 		\caption{The left plot shows the perturbation effect of $c_1$ when $c_2$ is fixed, while the right plot shows the perturbation effect of $c_2$ when $c_1$ is fixed for the second central moment of the birth-death process.}
 		\label{fig:BD_local2}
 	\end{center}
 \end{figure}  
 
 \begin{figure}[!htp]
 	\begin{center}
 		\includegraphics[width=.75\linewidth]{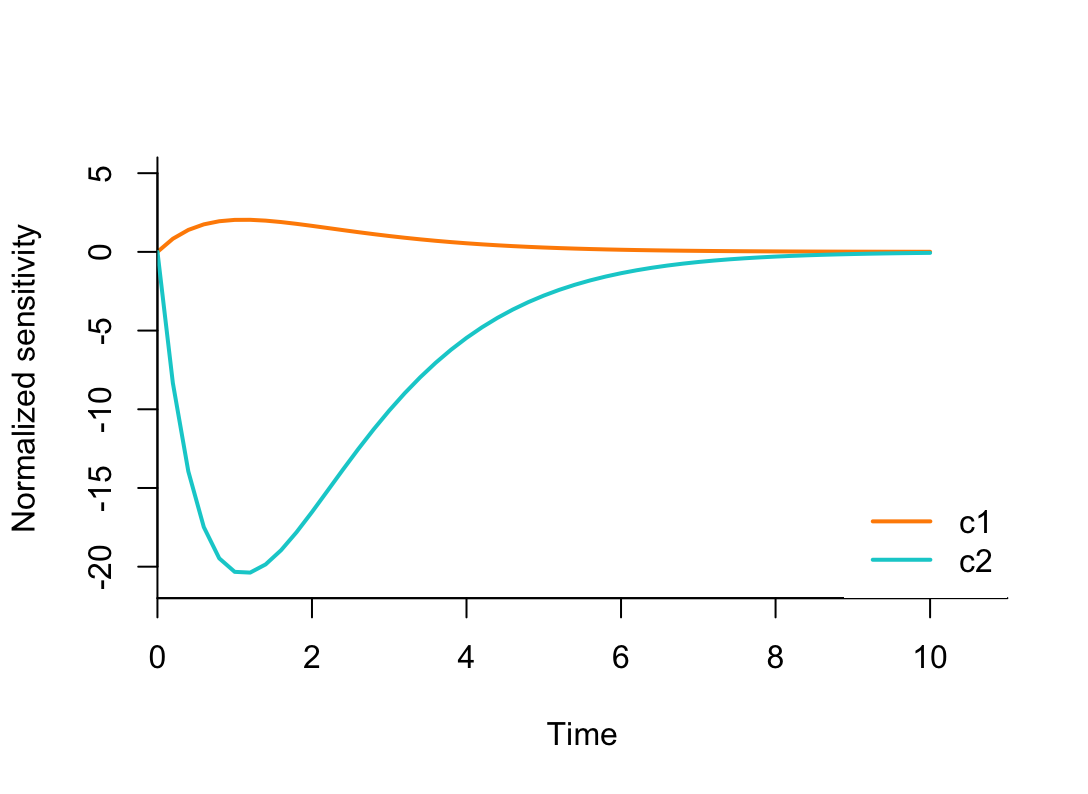}
 		\caption{Normalized sensitivity functions of the first moment to parameters $c_1$ and $c_2$ of the birth-death process.}
 		\label{fig:BD_localsens1}
 	\end{center}
 \end{figure}  
 
 \begin{figure}[!htp]
 	\begin{center}
 		\includegraphics[width=.75\linewidth]{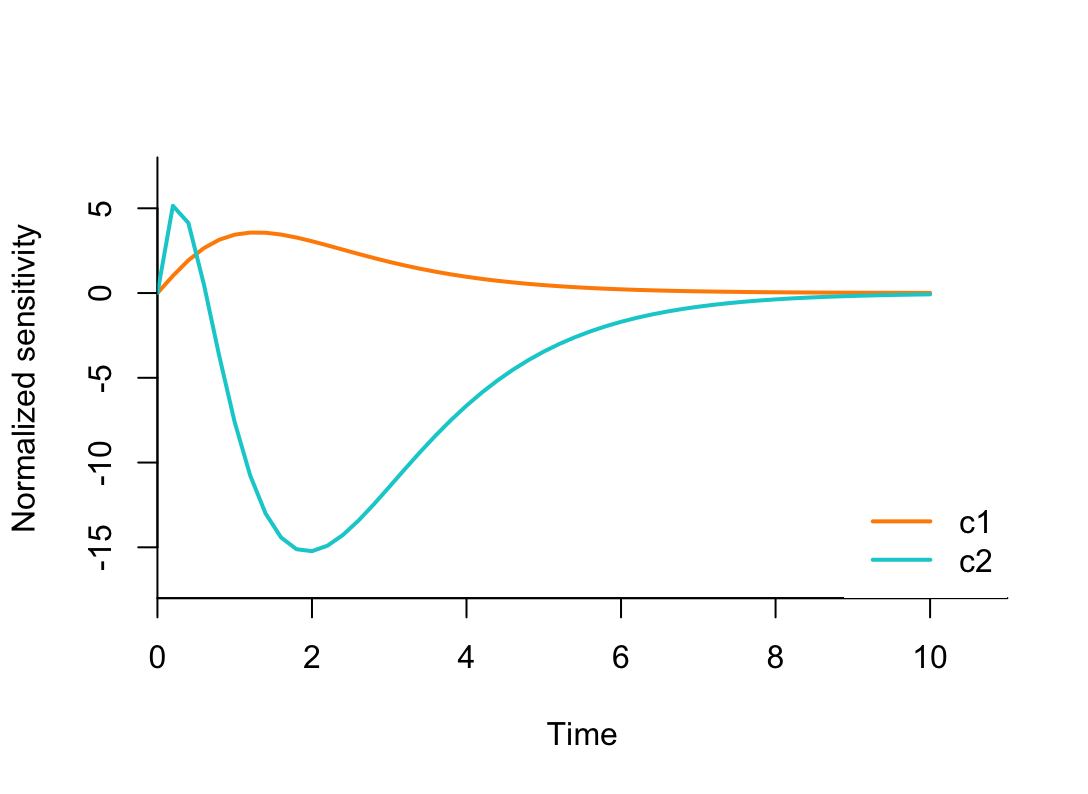}
 		\caption{Normalized sensitivity functions of the second central moment to parameters $c_1$ and $c_2$ of the birth-death process.}
 		\label{fig:BD_localsens2}
 	\end{center}
 \end{figure}  
 
 \begin{figure}[!htp]
 	\begin{center}
 		\includegraphics[width=.75\linewidth]{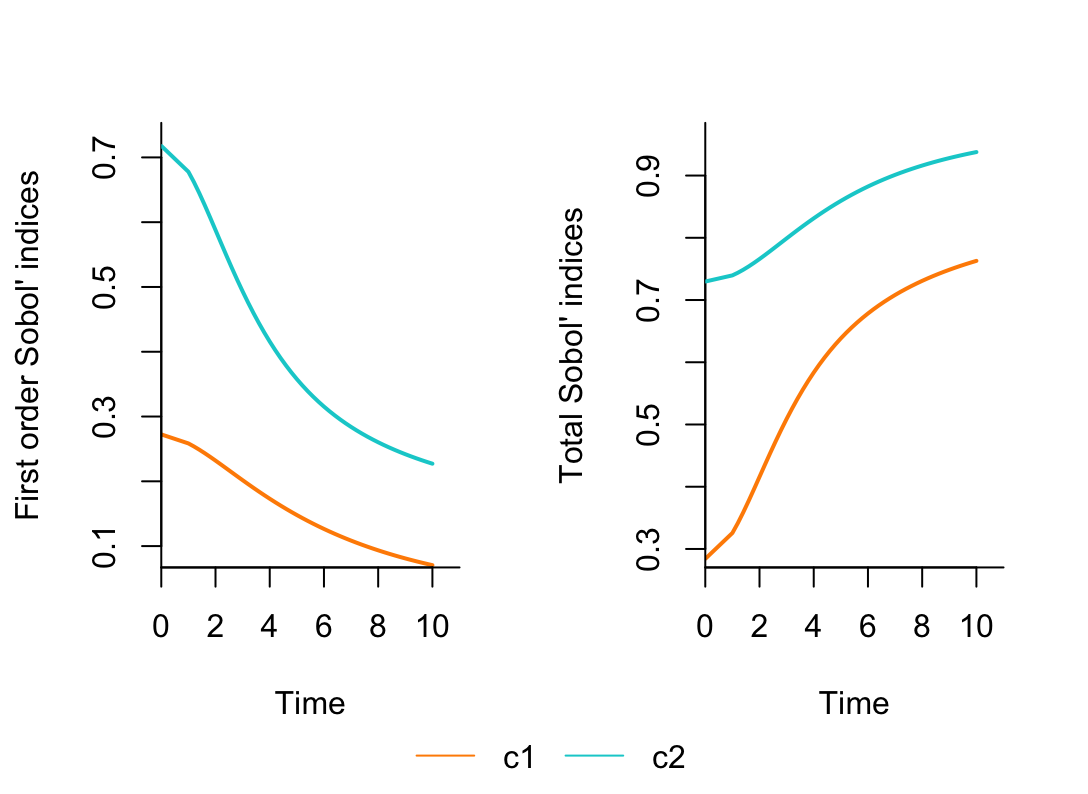}
 		\caption{Sobol' sensitivity indices for the first moment of the birth-death process.}
 		\label{fig:BD_global1}
 	\end{center}
 \end{figure}  
 
 \begin{figure}[!htp]
 	\begin{center}
 		\includegraphics[width=.75\linewidth]{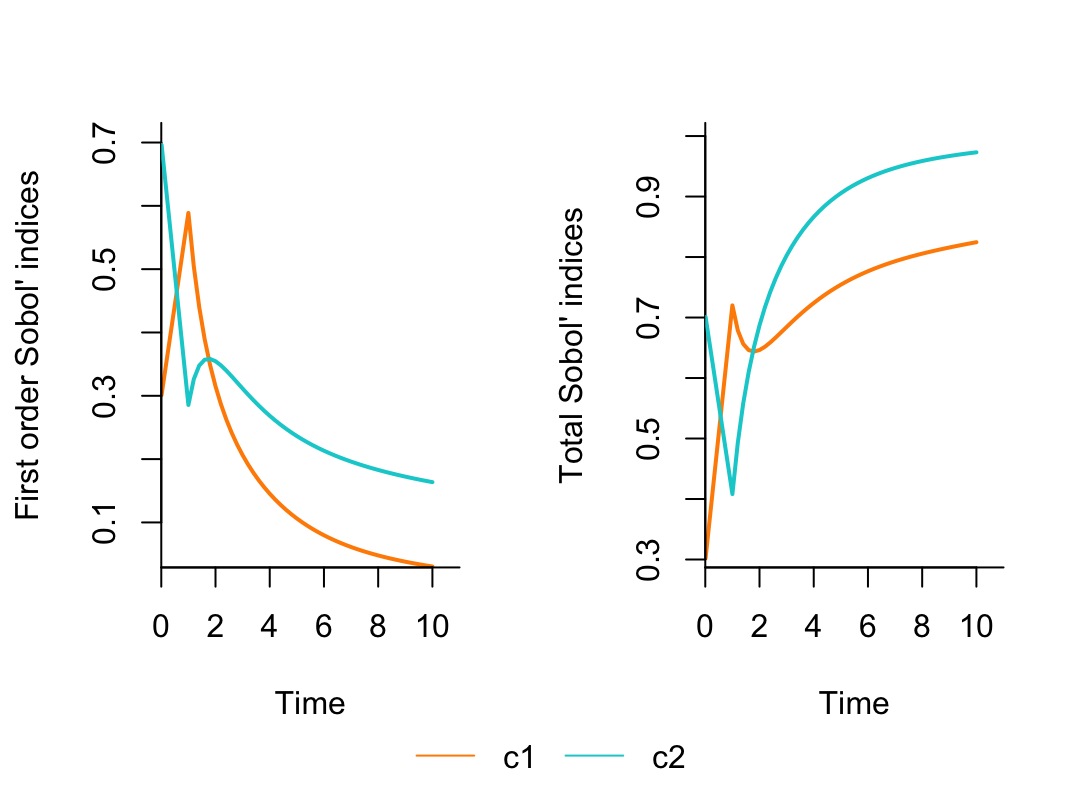}
 		\caption{Sobol' sensitivity indices for the second central moment of the birth-death process.}
 		\label{fig:BD_global2}
 	\end{center}
 \end{figure}  
 
 \subsubsection{Results and Discussions}
 In sensitivity analysis, the influence of parameter values on the model outcome is assessed. This analysis is conducted by modifying one of the parameter values and observing the resulting model output response. Since we divided our sensitivity analysis into two parts: local and global, we first analyze the results of local sensitivity. \\
 We begin by examining the effect of perturbing one parameter from its nominal or true value on the model's output. The most straightforward method to examine this change is graphically: increment each parameter by a small factor, such as 20\%, individually, execute the model with these perturbing values, and compare the results with the original output. For the birth-death process, we initially analyze the impact of each parameter separately on both the first and second central moments, utilizing the nominal or true values of the parameters ${\theta}_{true}= (c_1, c_2) = (0.10, 1.0)$ as a reference. In Figs. ~\ref{fig:BD_local1} and ~\ref{fig:BD_local2}, we observe the effects of perturbing each parameter while maintaining the other parameter fixed. Clearly, perturbing parameter $c_2$ significantly impacts both the first moment and the second central moment. In contrast, perturbing parameter $c_1$ results in only a minor effect on these moments. \\
 Next, we examine the normalized sensitivity of the model output concerning the parameter values through a series of sensitivity functions that are described in Eqs. ~\ref{eq:sens1} and ~\ref{eq:sens}. In our case, the first moment $\mu_1(t)$ and the second central moment $\sigma_{11}(t)$ are the output variables, where $c_1$ and $c_2$ are the parameters. For implementation the parameters are now perturbed by a factor of $10^{-8}$ and thus stay more closely in the vicinity of their nominal values. Both of the parameters are scaled with their nominal (true) values, while the output variables are not scaled. The greater the absolute sensitivity value, the more significant the parameter; that is, when the absolute sensitivity value is larger, the parameter is considered to have more importance. The graph of these sensitivity functions shows how they vary with time in Figs. ~\ref{fig:BD_localsens1} and ~\ref{fig:BD_localsens2}. From Fig. ~\ref{fig:BD_localsens1}, we can see that the sensitivity function for the first moment is always positive for parameter $c_1$, while it is negative for parameter $c_2$ and shows no effect after the time $t=8s$. Similarly, Fig. ~\ref{fig:BD_localsens2} shows that parameter $c_2$ is more sensitive to the sensitivity function of the second central moment than parameter $c_1$. \\
We now analyze the global sensitivities of the birth-death process using Sobol' method. Each of the state variables $\mu_1(t)$ and $\sigma_{11}(t)$ corresponds to the model function $f$ from subsection ~\ref{sec:sobol}. The parameters $c_1$ and $c_2$ are regarded as the input variables for the Sobol' sensitivity analysis. Therefore, we will examine the sensitivity of the first moment and the second central moment with regard to changes in these two parameters. Since we are implementing the Sobol’-Martinez method, we choose $n = 15000$ simulations for Monte Carlo estimations. We assume a uniform distribution for all parameters but with different lower and upper boundaries given in Table ~\ref{table:BD_bound}. Figs. ~\ref{fig:BD_global1} and ~\ref{fig:BD_global2} showed the first-order Sobol' indices and total Sobol' indices for the first moment and second central moment, respectively. Since there are only two parameters, the total effect of parameter $c_1$, denoted as $S_{T_1}=S_1+S_{12}$, is defined such that $S_1$ represents the direct effect of $c_1$, while $S_{12}$ indicates the interaction effect between $c_1$ and $c_2$. In a similar manner, the total effect of parameter $c_2$, denoted as $S_{T_2}=S_2+S_{21}$, is defined such that $S_2$ represents the direct effect of $c_2$, while $S_{21}$ indicates the interaction effect between $c_2$ and $c_1$. For the interpretation, we will focus on the total Sobol’ indices since they present us more robust information when there is interaction between parameters. As can be seen from Fig. ~\ref{fig:BD_global1}, the parameter $c_2$ has the largest overall influence on the first moment $\mu_1(t)$ for nearly the whole time from $0$ to $10$ seconds, followed by parameter $c_1$. For the second central moment $\sigma_{11}(t)$, the influence of parameter $c_2$ decreases for a time between $0$ to $2$ seconds, and thereafter, it exceeds the overall influence of parameter $c_1$. For all the local and global sensitivity analysis simulations for the birth-death process, we choose true parameter values: ${\bm{\theta}_{true}}= (c_1, c_2) = (0.10, 1.0) $ and initial conditions $(\mu_1(0), \sigma_{11}(0)) = (50, 0)$ and time $t=10s$.

\begin{figure}[!htp]
	\begin{center}
		\includegraphics[width=.75\linewidth]{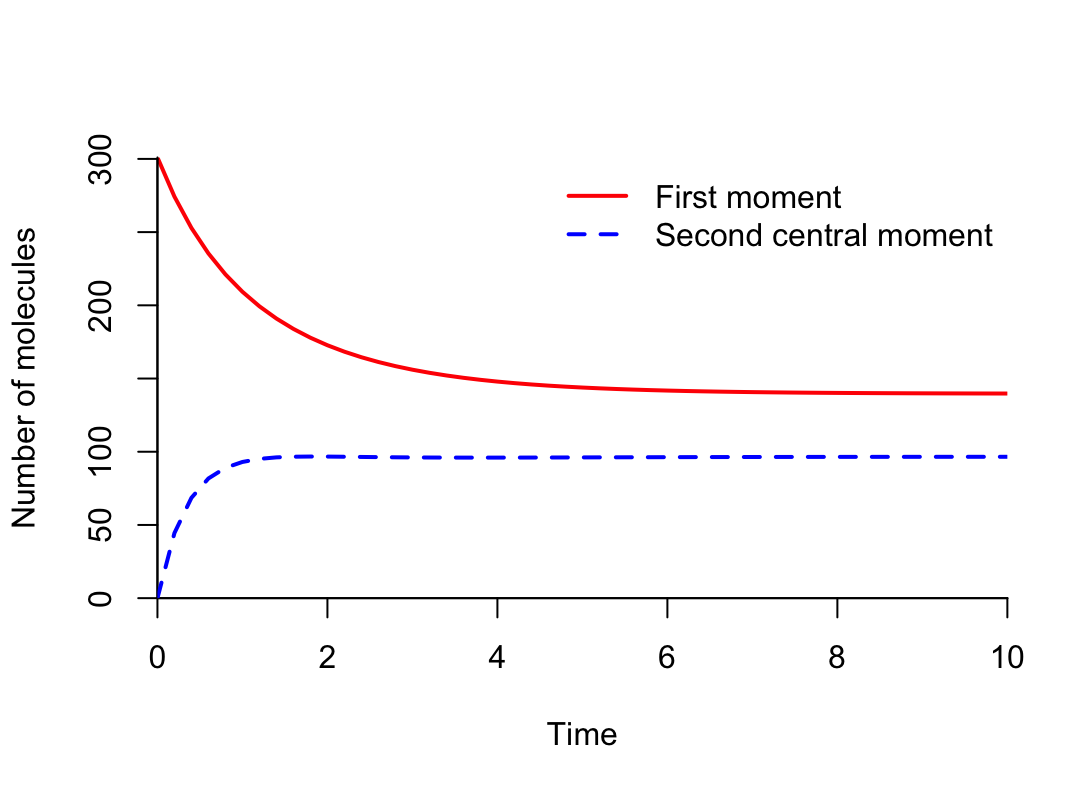}
		\caption{First moment and second central moment of the dimerization process.}
		\label{fig:DM}
	\end{center}
\end{figure}

 \subsection{Dimerization Process} 
 The reversible dimerization process, which consists of two molecular species, namely dimer $X$ and monomer $Y$, has the following reactions:
 $$
 \begin{array}{cc} 
 	R_1: & 2X \stackrel{c_1}{\longrightarrow}  Y \\ [10 pt]
 	R_2: &  Y \stackrel{c_2}{\longrightarrow} 2X
 \end{array}
 $$
 where in reaction $R_{1}$ two monomers $X$ fuse into a single dimer $Y$ and in reaction $R_{2}$ the dimer split apart into two monomers $X$. The propensity functions for $R_1$ and $R_2$ are given by $\alpha_1 = c_1 \frac{x(x-1)}{2}$ and $\alpha_2 = c_2 \frac{(x_{0}-x)}{2}$, respectively, where the associated stoichiometric matrix is 
 $$
 S=
 \begin{pmatrix}
 	-2 & 1 \\
 	2 & -1 
 \end{pmatrix}
 $$
 By using Eq.~\eqref{eq:CME}, we can find the equation of CME for the dimerization process\cite[p.~55]{gillespie2009moment} as follows:
 
 \begin{equation}
 	\label{eq:Dimer_CME}
 	\begin{aligned}
 		\frac{d P(x, t)} {d t}=  \frac {P(x+2, t) c_1(x+2)(x+1)}{2}  +\frac {P(x-2, t) c_2\left(x_0-x+2\right)}{2} 
 		- \frac {P (x, t) c_1 x(x-1)} {2} +c_2 \frac {(x_0-x)}{2} 
 	\end{aligned}
 \end{equation}
Since the analytical solution as well as the direct numerical integration of Eq.~\eqref{eq:Dimer_CME} is intractable, we simulate the dimerization process using zero moment approximation. By utilizing Eqs.~\eqref{eq:CME_MOM_initiial5} and ~\eqref{eq:CME_MOM_initiial6}, we can derive the following moment-based ODEs\cite[Supporting Information S1 p.~4]{kugler2012moment}: 
 \begin{equation}
 	\label{eq:dimer_mean_fun}
 	\begin{aligned}
 		\frac{d \mu_{1} (t)}{d t}= c_1 \mu_{1}(t)\left(1-\mu_{1}(t)\right)+
 		c_2\left(x_0-\mu_{1}(t)\right)-c_1 \sigma_{11}(t), \mu_1(0)=x_0 
 	\end{aligned}
 \end{equation}
 \begin{equation} 
 	\label{eq:dimer_var_fun}
 	\begin{aligned}
 		\frac{d \sigma_{11} (t)}{d t}= - 2 c_1\left(2 \mu_{1}(t)+2\right) \sigma_{11}(t)-2 c_2 \sigma_{ 11} (t)
 		+2 c_1 \mu_{1}(t)\left(\mu_{1}(t)-1\right) 
 		+2 c_2\left(\mathbb{X}_{0}-\mu_{1}(t)\right),  \sigma_{11}(0)=0  
 	\end{aligned}
 \end{equation}
 where $\mu_1$ is the first moment and $\sigma_{11}$ is the second central moment of the continuous state variable ${\mathbb{X}}(t)$ of the monomers. The solution of the first moment and second central moments are depicted in Fig.~\ref{fig:DM} with the known (true) parameter values: ${\bm{\theta}_{true}}= (c_1, c_2) = (1.66\times 10^{-3}, 0.2)$ and initial conditions $(\mu_1(0), \sigma_{11}(0)) = (301, 0)$\cite{ale2013general} for the time $t=10s$.
 
  \begin{figure}
 	\begin{center}
 		\includegraphics[width=.75\linewidth]{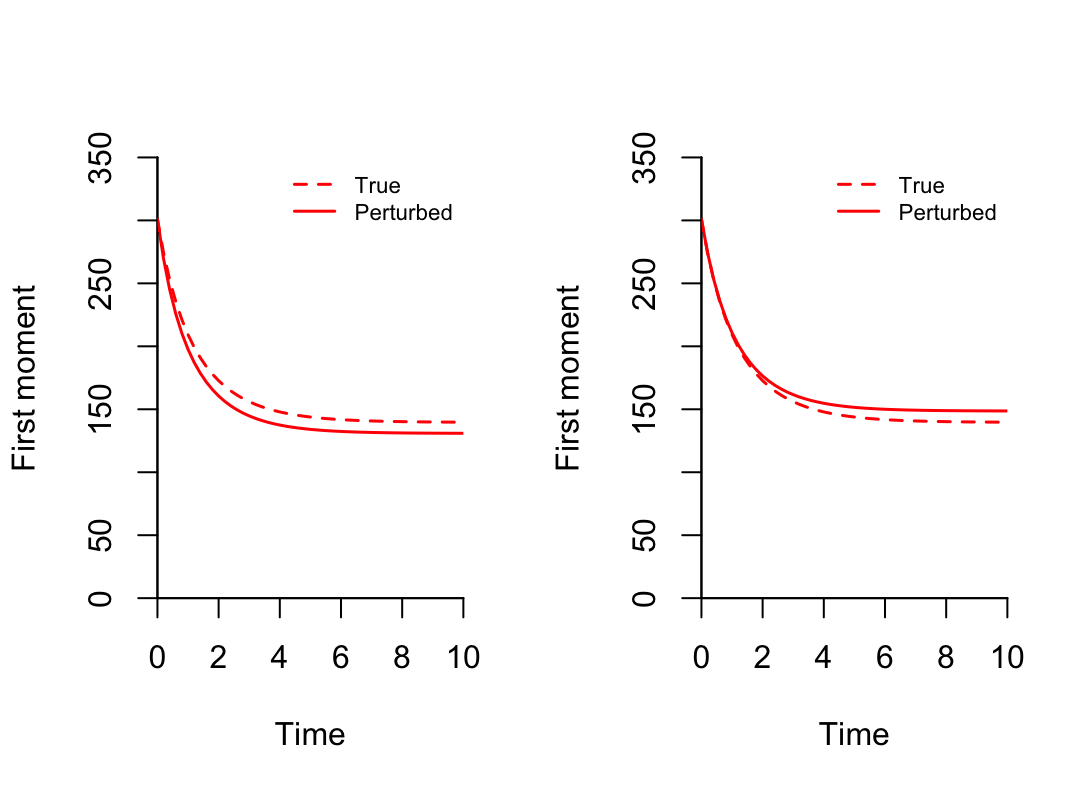}
 		\caption{The left plot shows the perturbation effect of $c_1$ when $c_2$ is fixed, while the right plot shows the perturbation effect of $c_2$ when $c_1$ is fixed for the first moment of the dimerization process.}
 		\label{fig:DM_local1}
 	\end{center}
 \end{figure}  
 
 \begin{figure}
 	\begin{center}
 		\includegraphics[width=.75\linewidth]{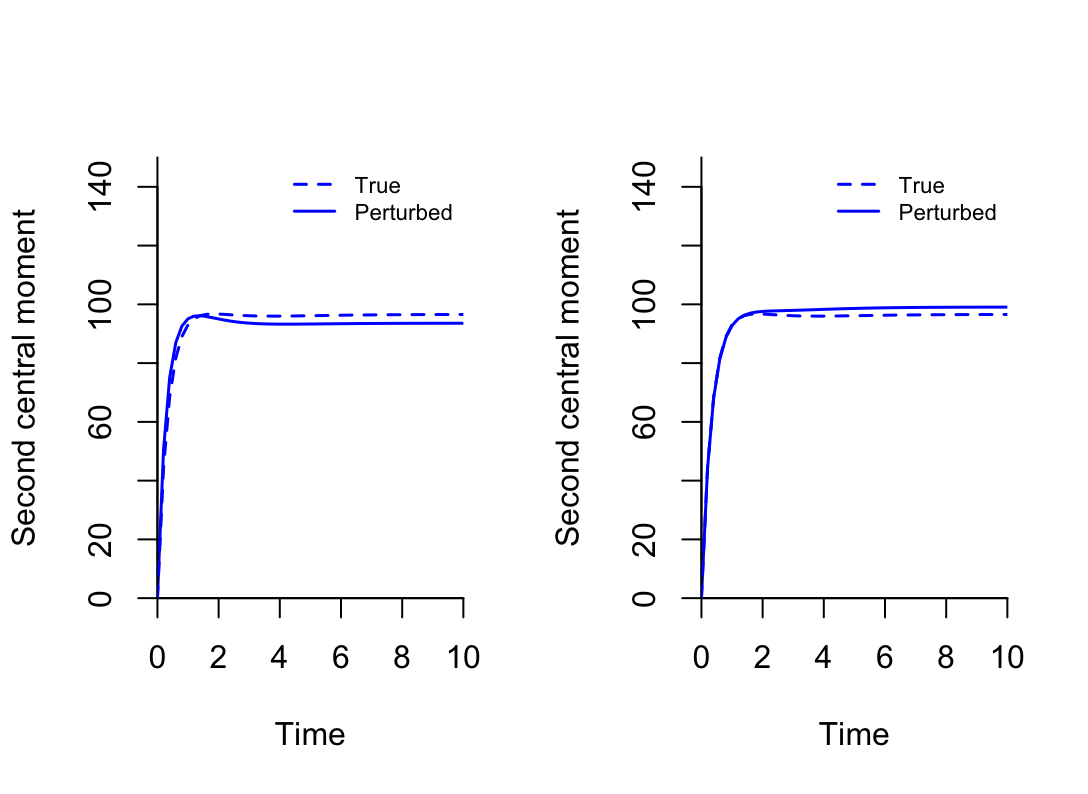}
 		\caption{The left plot shows the perturbation effect of $c_1$ when $c_2$ is fixed, while the right plot shows the perturbation effect of $c_2$ when $c_1$ is fixed for the second central moment of the dimerization process.}
 		\label{fig:DM_local2}
 	\end{center}
 \end{figure}  
 
 \begin{figure}
 	\begin{center}
 		\includegraphics[width=.75\linewidth]{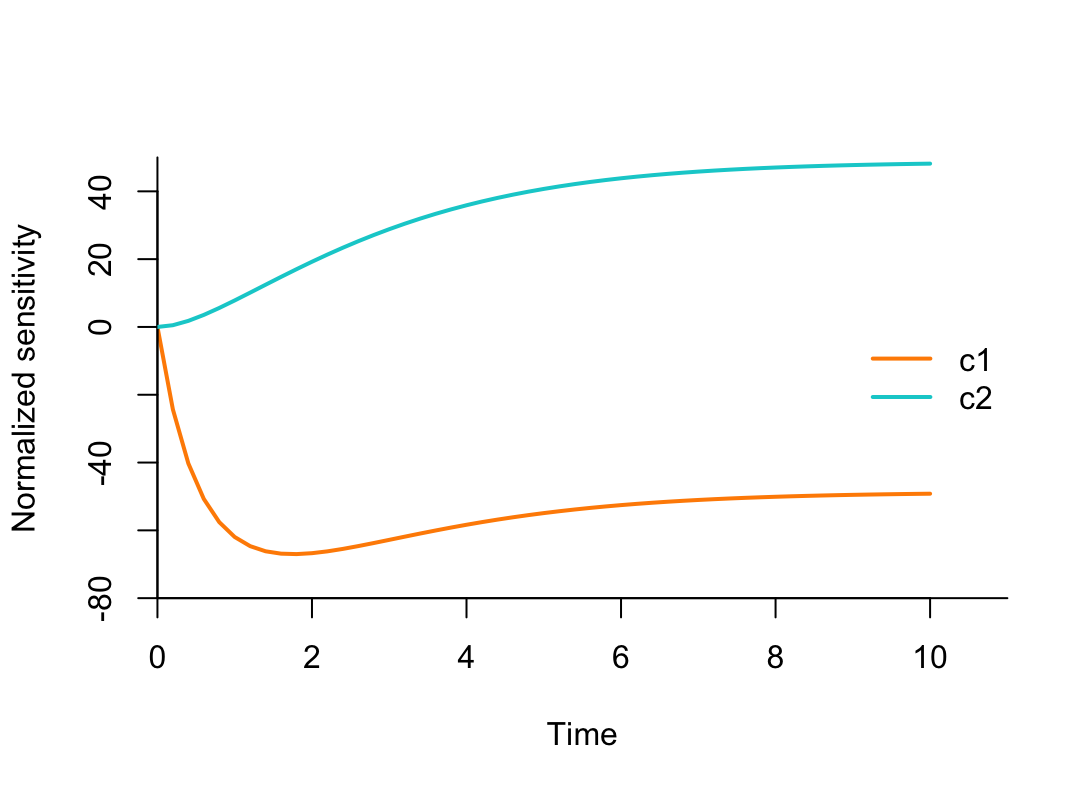}
 		\caption{Normalized sensitivity functions of the first moment to parameters $c_1$ and $c_2$ of the dimerization process.}
 		\label{fig:DM_localsens1}
 	\end{center}
 \end{figure}  
 
 \begin{figure}
 	\begin{center}
 		\includegraphics[width=.75\linewidth]{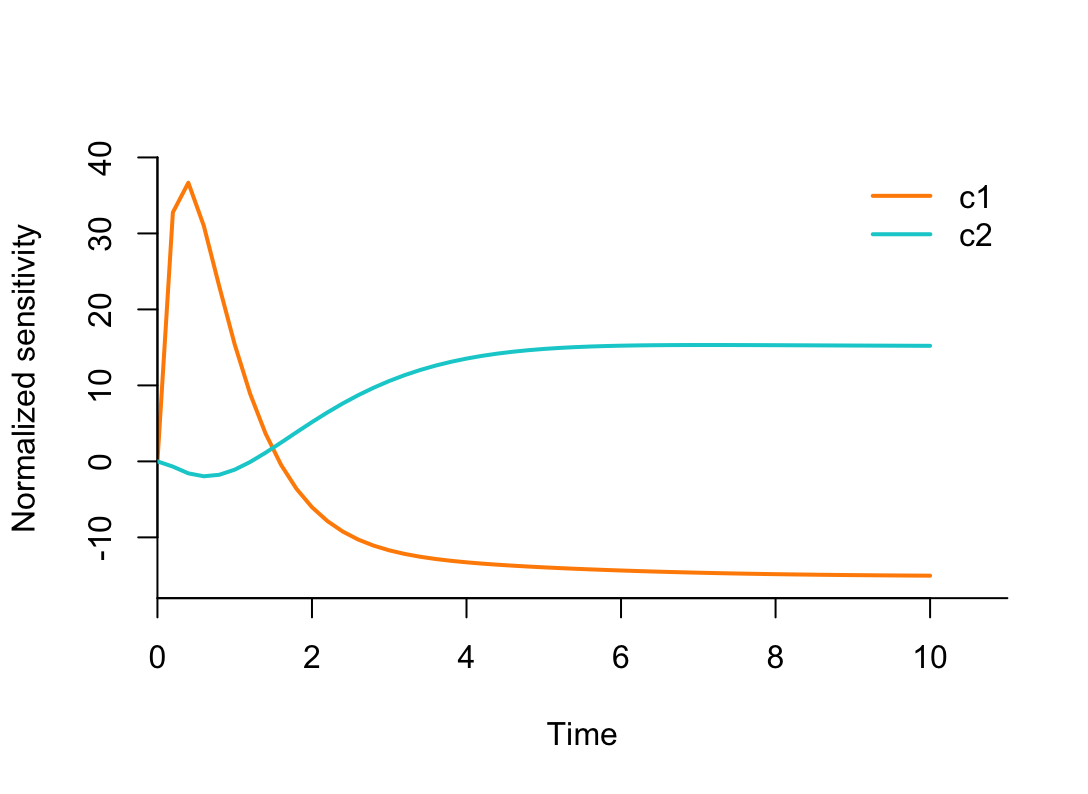}
 		\caption{Normalized sensitivity functions of the second central moment to parameters $c_1$ and $c_2$ of the dimerization process.}
 		\label{fig:DM_localsens2}
 	\end{center}
 \end{figure}

 \begin{figure}
 	\begin{center}
 		\includegraphics[width=.75\linewidth]{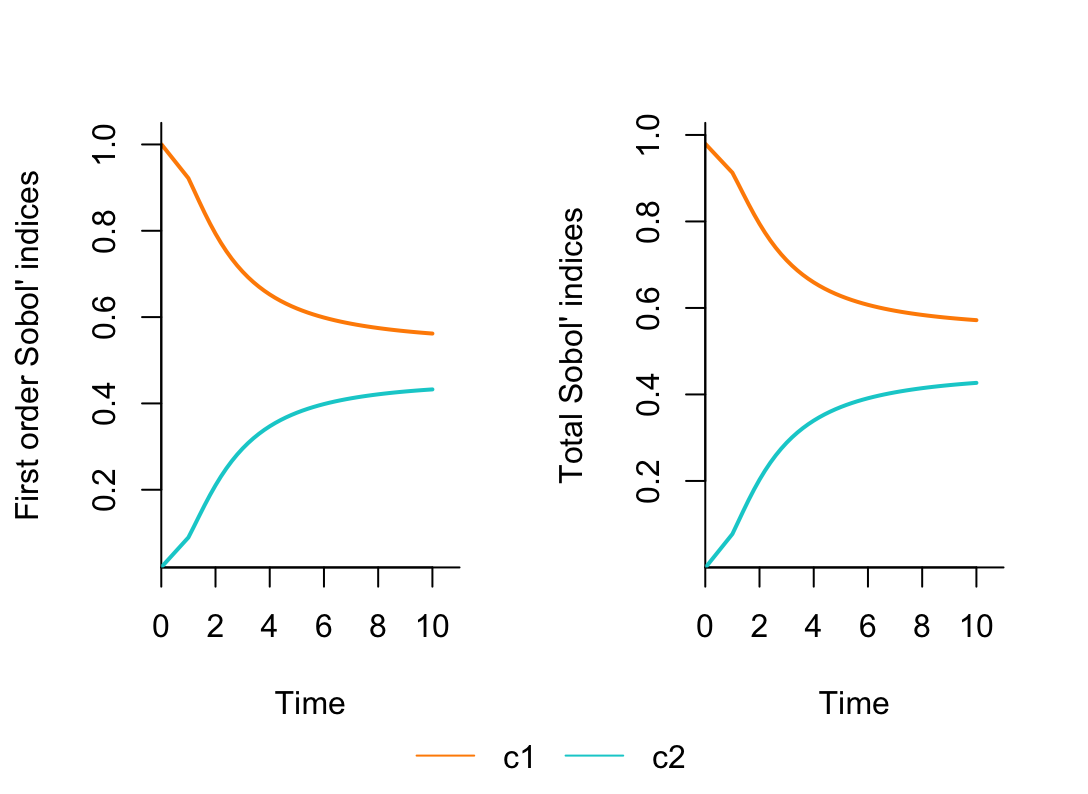}
 		\caption{Sobol' sensitivity indices for the first moment of the dimerization process.}
 		\label{fig:DM_global1}
 	\end{center}
 \end{figure} 
 
 \begin{figure}
 	\begin{center}
 		\includegraphics[width=.75\linewidth]{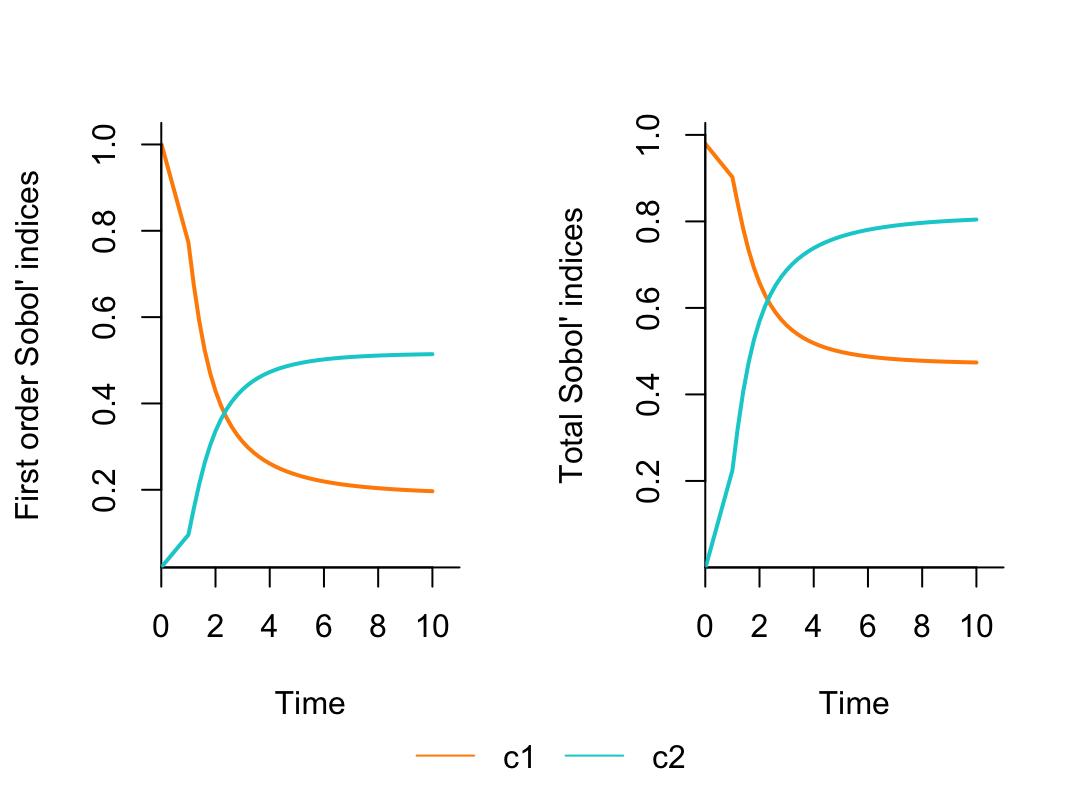}
 		\caption{Sobol' sensitivity indices for the second central moment of the dimerization process.}
 		\label{fig:DM_global2}
 	\end{center}
 \end{figure} 
 
  \subsubsection{Results and Discussions}
For local sensitivity analysis of the dimerization process, we perturb each parameter by 20\%, individually, execute the model with these perturbing values, and compare the results with the original output. In Figs. ~\ref{fig:DM_local1} and ~\ref{fig:DM_local2}, we observe the effects of perturbing each parameter while maintaining the other parameter fixed. We see that perturbing parameter $c_1$ has a slightly larger impact on both the first moment and the second central moment compared to the parameter $c_2$. The normalized sensitivity of the model output is shown in Figs. ~\ref{fig:DM_localsens1} and ~\ref{fig:DM_localsens2}. Here we perturb both parameters $c_1$ and $c_2$ by a factor of $10^{-8}$ and scale them with their nominal (true) values, while the output variables are not scaled. From Fig. ~\ref{fig:DM_localsens1}, we can see that the sensitivity function for the first moment is always positive for parameter $c_2$, while it is negative for parameter $c_1$ and has a large effect on the first moment $\mu_1(t)$. Similarly, Fig. ~\ref{fig:DM_localsens2} shows that parameter $c_1$ is more sensitive to the sensitivity function of the second central moment $\sigma_{11}(t)$ than parameter $c_1$. 
  
   \begingroup
  \setlength{\tabcolsep}{9pt} % Default value: 6pt
  \renewcommand{\arraystretch}{1.2} 
  \begin{table}[h]
  	\caption{Bounds on the support of the uniform distribution of the parameters in the dimerization process.}
  	\begin{center}
  		\begin{tabular}{@{}c c c @{}} 
  			\hline
  			parameter & lower bound & upper bound  \\
  			\hline 
  			$c_{1}$  & $1.0\times 10^{-4}$ & $9.0\times 10^{-3}$ \\
  			$c_{2}$  & $0.01$ & $1.0$ \\
  			\hline
  		\end{tabular}
  	\end{center}
  	\label{table:DM_bound}
  \end{table}
  \endgroup 
  
  {\noindent For global sensitivity analysis of the dimerization process using the Sobol' method, parameters $c_1$ and $c_2$ are considered input variables. Here we again use the Sobol’-Martinez method, where $n = 15000$ simulations are chosen for Monte Carlo estimations. We assume a uniform distribution for all parameters but with different lower and upper boundaries given in Table ~\ref{table:DM_bound}. Figs. ~\ref{fig:DM_global1} and ~\ref{fig:DM_global2} showed the first-order Sobol' indices and total Sobol' indices for the first moment and second central moment, respectively. The total effects of parameters $c_1$ and $c_2$ are denoted as $S_{T_1}$ and $S_{T_2}$, respectively, and they follow the same formulation as in the birth-death process. The total Sobol's indices can be explained, as seen in Fig. ~\ref{fig:DM_global1}, the parameter $c_1$ has the largest overall influence on the first moment $\mu_1(t)$ for nearly the whole time from $0$ to $10$ seconds. For the second central moment, $\sigma_{11}(t)$, in Fig. ~\ref{fig:DM_global2}, the influence of parameter $c_1$ decreases for a time between $0$ and $10$ seconds, and the overall influence of parameter $c_2$ increases for the time between $0$ and $10$ seconds. For all the local and global sensitivity analysis simulations for the dimerization process, we choose true parameter values: ${\bm{\theta}_{true}}= (c_1, c_2) = (1.66\times 10^{-3}, 0.2) $ and initial conditions $(\mu_1(0), \sigma_{11}(0)) = (301, 0)$ and time $t=10s$.}
  
 \section{Conclusion}
 \label{sec:Conclusion}
  Accurately and effectively estimating sensitivities in stochastic biochemical models continues to be a significant challenge. For biochemical reaction networks, where stochastic effects are crucial, the moment-based method is attractive since the integration of the moment equations is frequently quick in comparison to alternative approaches such as the stochastic simulation algorithm for solving the chemical master equation. This paper provides the framework for the local and global sensitivity analysis with a focus on using the moment-based method in stochastic biochemical models. Even though different sensitivity analysis approaches are available for the dynamic analysis of models, we used perturbation and forward finite difference approximations for local sensitivity analysis and Sobol's method for global sensitivity analysis. In general, sensitivity analysis is a useful starting point for identifying the input parameters that have a significant impact on the output behaviors of the model. However, it is important to exercise caution regarding the precision of sensitivity analysis results, particularly when model parameters lack identifiability. The results of a local sensitivity analysis may be over-interpreted when predicting the knockout consequence of corresponding biochemical reactions, as local sensitivity relies on small parameter perturbations.

\bibliographystyle{unsrt}
\bibliography{bibliography}

\end{document}